\documentclass[aps, prl, twocolumn,showpacs,preprintnumbers,amsmath,amssymb]{revtex4}
\usepackage{graphicx}
\begin{document}
\title{Search for stable Strange Quark Matter in lunar soil}

\author{K. Han$^1$}\author{J. Ashenfelter$^2$}
\author{A. Chikanian$^1$}\author{W. Emmet$^1$}
\author{L. E. Finch$^1$}
\author{A. Heinz$^{2}$}
\author{J. Madsen$^3$}
\author{R. D. Majka$^1$}\author{B. Monreal$^4$\footnote{ Current address:
Department of Physics, University of California, Santa Barbara, CA 93106, USA}}
\author{J. Sandweiss$^1$}
\affiliation{%
$^1$Physics Department, Yale University, New Haven, Connecticut 06520, USA
\\$^2$A. W. Wright Nuclear Structure Laboratory, Yale University, New Haven, Connecticut 06520, USA
\\$^3$Department of Physics and Astronomy, University of Aarhus, DK-8000 {\AA}rhus C, Denmark
\\$^4$Laboratory for Nuclear Science, Massachusetts Institute of Technology, Cambridge, Massachusetts 02139, USA}
\date{\today}

\begin{abstract}
We report results from a search for strangelets (small chunks of
Strange Quark Matter) in lunar soil using the Yale WNSL accelerator
as a mass spectrometer.  We have searched over a range in mass from
A=42 to A=70 amu for nuclear charges 5, 6, 8, 9, and 11. No
strangelets were found in the experiment. For strangelets with
nuclear charge 8, a concentration in lunar soil higher than
$10^{-16}$ is excluded at the 95\% confidence level. The implied
limit on  the strangelet flux in cosmic rays is the most sensitive
to date for the covered range and is relevant to both recent
theoretical flux predictions and a strangelet candidate event found
by the AMS-01 experiment.
\end{abstract}
\maketitle

Strange Quark Matter (SQM) is a proposed state of hadronic matter
made up of roughly one-third each of up, down, and strange quarks in
a single hadronic bag that can be as small as an atomic nucleus or
as large as a star. It has been over 30 years since the first
suggestion that the true ground state of cold hadronic matter might
be SQM rather than nuclear
matter~\cite{Bodmer:1971we,Witten:1984rs}. If true, the implications
would be tremendous for both basic research and applied
science~\cite{Shaw:1988pc}. With this motivation, many searches for
stable SQM have been undertaken using a variety of methods. These
searches have collectively observed a handful of interesting events
but have neither been able to find compelling evidence for stable
SQM nor to rule out its existence.

The idea that Quark Matter made of only up and down quarks is stable
can be dismissed immediately by the observation that normal nuclear
matter does not decay into it.  However, in the case of SQM such a
decay would require many simultaneous weak interactions, making it
prohibitively unlikely.  The stability of SQM cannot yet be
determined from first principles within QCD, but has been addressed
in various phenomenological models.  The most commonly used of these
is the MIT Bag Model~\cite{Chodos:1974je,Farhi:1984qu}, which also
has been extended to include the effects of color flavor locking
(CFL)~\cite{ Alford:1998mk,Madsen:2001fu}. The results of such
calculations are inconclusive, but for a significant part of the
reasonable parameter space in these models, SQM is in fact
absolutely stable for baryon number greater than some minimum value,
$A_{min}$~\cite{ Farhi:1984qu,Madsen:1993PRL}. $A_{min}$ is
typically found to be larger than 50 and smaller than 1000 although
shell effects which are important for $A \lesssim 100$ may cause
islands of stability at $A$ values smaller  than $A_{min}$. The key
point is that SQM stability is a question that must be settled
experimentally or observationally.

If SQM is stable at zero pressure, all compact stars which are
commonly thought of as neutron stars may in fact be "strange stars",
i.e. composed of SQM~\cite{Page:2006ud}. A strange star which is a
member of a binary system will eventually suffer a collision with
its partner, possibly resulting in the ejection of some fraction of
its mass in the form of strangelets. This should ultimately lead to
a flux of strangelets in cosmic rays ~\cite{Madsen:2004vw,
MedinaTanco:1996je}. The resulting flux of strangelets at the Moon
is estimated to be about 2000 per (m$^2$ year
sterad)~\cite{Madsen:2004vw} assuming all the ejected SQM mass were
in the form of strangelets of one particular baryon number (and is a
lower limit to the integrated flux if the SQM mass is distributed
below a given $A$), which over-simplifies the real scenario. It is
not, however, unreasonable to expect that the distribution may be
clustered broadly around $A_{min}$~\cite{KrishnaNew} so that near
$A_{min}$ this flux estimation may not be a gross overestimation.
Recent strange star collision simulations~\cite{Bauswein:2008gx}
show that the calculation in Ref.~\cite{Madsen:2004vw} may
underestimate the total galactic ejection rate by one or two orders
of magnitude for strongly bound SQM, whereas the ejection rate may
be negligible for loosely bound SQM. Given the large uncertainties,
the theoretical calculation~\cite{Madsen:2004vw} should be
considered a very rough guide.

Previous experiments (reviewed
in~\cite{Klingenberg:1999sb,Finch:2006pq}) have searched for cosmic
strangelet relics in terrestrial materials, meteorites, and lunar
soil. There have also been satellite and balloon-borne detectors
which would be sensitive to a possible strangelet component in
cosmic radiation. Some searches for strangelets with particular
nuclear charges have reported negative results at sensitivity levels
lower than theoretical predictions~\cite{Finch:2006pq}. While these
results rule out certain strangelet charge states at this level,
they do not generally disprove the hypothesis of stable SQM or
strange stars.

Meanwhile, candidate events consistent with strangelet
characteristics have been published by several
experiments~\cite{Price_ET,Saito_ET,Ichimura_ET}. The search
reported here was specifically motivated by two SQM candidate events
found by the Alpha Magnetic Spectrometer (AMS) collaboration during
the AMS-01 prototype flight in 1998~\cite{Aguilar:2002ad}.  One of
the events was reconstructed as having a nuclear charge of $Z=+8$
and a mass $A=54^{+8}_{-6}$, which we will denote as
$^{54}\mathrm{O}$. These two events were not published in
anticipation of the mounting of the full AMS experiment (AMS-02) on
the International Space Station (ISS) which was originally scheduled
for 2003 and would easily prove or disprove these events.  With the
delay of the AMS experiment, it has become interesting to follow up
on these events by other means.

In this paper we report a search for low mass ($A \approx 54$)
strangelet relics in lunar soil using the A. W. Wright Nuclear
Structure Laboratory (WNSL) tandem Van de Graaff
accelerator~\cite{WNSL_Tandem88} as a mass spectrometer. The
advantage of using lunar soil over using terrestrial material for
such a study is that the Moon has neither magnetic field (so low
energy strangelets are not turned away) nor geological activity (so
that strangelets that stop near the surface tend to remain there for
hundreds of millions of years). The search reaches single event
sensitivity levels around 3 parts in $10^{17}$ and the implied
sensitivity to SQM as a component of cosmic rays falls near the
theoretical flux prediction and below the flux implied by the AMS-01
candidate event.

\begin{figure}
\includegraphics[height=0.45\textwidth, angle=90]{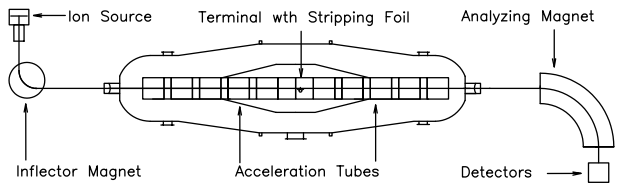}
\caption{\label{fig:apparatus} Schematic figure of the Yale WNSL Tandem
accelerator (not to scale). Beam direction is from left to right.}
\end{figure}

For this experiment, we obtained 15 g of lunar soil sample No. 10084
from NASA. This fine particulate sample was collected from the top
7.5 cm of the lunar surface~\cite{Apollo11PSR} and has a cosmic ray
exposure age of $520\pm 120$ Myr~\cite{SoilAge}. This sample is used
(in 0.1 g increments) as source material for the tandem accelerator.

The Yale WNSL accelerator is shown schematically in
Fig.~\ref{fig:apparatus}. Negative ions of the source material are
formed in a cesium sputter source~\cite{IonSource_Middleton83} and
accelerated to 20~keV before undergoing a $90^\circ$ bend in the
inflector magnet. The inflector magnet is set to transmit ions with
a given mass $A_0$ and charge $Q=-1$. Following this, the ions enter
the main acceleration tank and are accelerated towards the positive
terminal at 17~MV, where a 10 $\mathrm{\mu g / cm^2}$ carbon foil
strips electrons from the ions.  For 17~MeV $^{54}\mathrm{O}$
strangelets, the most probable stripped charge state is
$Q=+5$~\cite{Shima:1986}. The stripped (and therefore positive) ions
are accelerated away from the terminal to ground, and then go
through another $90^\circ$ bend in the analyzing magnet, which is
set to transmit charge $Q=+5$ (total energy 102~MeV) ions only. The
mass acceptance $\delta_m / m$ of the inflector and analyzing
magnets (with all slits wide open) are both approximately 0.6\% so
that a mass range of $\delta_m = 1/4$ amu can be covered in each
run. Finally, the ions enter our detector system as described below.

The long and short term performance of the accelerator was observed
closely. Electrostatic accelerators usually rely on ion beam
feedback to regulate their terminal voltages. However, in our
experiment, when set for a mass $A_0$ which is not an integer, there
is no normal nuclear ion beam transmitted through the machine.
Therefore, the terminal voltage is held constant by a feedback
system utilizing a set of Generating Voltmeters inside the
accelerator tank wall. The short term ($\sim1$ hour) stability of
the accelerator control system was verified with known beams and
then monitored for strangelet runs by observing the beam current in
a Faraday cup near our detector system and the beam position on a
ZnS fluorescent screen. The long term stability was monitored by
periodic short checks, performed roughly every 4 hours, of the
transmission of beams of known elements within (or doped into) the
lunar soil and readjustment of the terminal voltage when
appropriate.

When set for a mass $A_0$ different from any normal nuclear mass,
the accelerator and beam transport itself gives a background
rejection for a strangelet search on the order of 1 part in
$10^{12}$.  For integer values of $A_0$, the rejection can be as
poor as $10^{-3}$. To reach a level of $10^{-17}$ over the entire
mass range, we use a detector system after the analyzing magnet.

Because strangelets are expected to have nearly as many strange
quarks as up and down quarks, a commonly used experimental signature
for strangelets is a much smaller nuclear charge to mass ratio than
normal nuclei.  In this experiment, we exploit the fact that
strangelets' charge-to-mass ratio gives them a smaller dE/dx and a
larger stopping range than normal nuclei of similar mass and
incident energy.

\begin{figure}
\includegraphics[height=0.4\textwidth, angle=270]{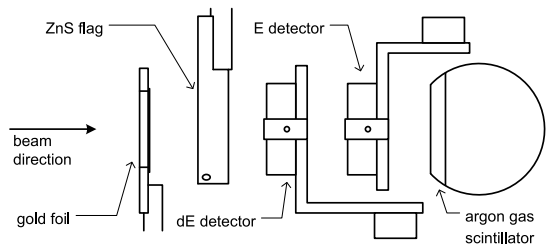}
\caption{\label{fig:detectors} Schematic setup of the detector
system. All the components, except the scintillator, can be
withdrawn from and inserted into the beam line remotely.}
\end{figure}

Our detector system is shown schematically in
Fig.~\ref{fig:detectors}. All the components except the argon
scintillator can be withdrawn from or inserted into the beam line
remotely. The argon gas scintillation counter and the ZnS screen
(viewed by a camera imaging the screen) were used as monitors of
beam quality and stability throughout the running period and to make
various transmission measurements.

For normal running during the strangelet search, the gold foil and
both silicon counters are put into the beam line. The Au foil
thickness of 10 $\mathrm{\mu m}$ is chosen so that a 102 MeV
strangelet $^{54}\mathrm{O}$ entering the foil will exit with about
40 MeV.  A normal nucleus of comparable mass and incident energy
will stop in a length of 7 $\mathrm{\mu m}$.  To penetrate the foil,
an ion must have higher energy and/or less mass (and so should
generally be inconsistent with rigidity selection through the
accelerator).

The final level of background discrimination comes from our silicon
(dE, E) telescope. The first silicon (dE) detector, which measures
the energy loss of a penetrating strangelet, has a circular
cross-sectional active area of 40 $\mathrm{mm^2}$ and is $11.7~\pm
1.3\, \mathrm{\mu m}$ thick.  The second silicon (E) detector, which
measures the remaining energy, has a cross section of
100~$\mathrm{mm^2}$ and is 100~$\mathrm{\mu m}$ thick.  The energy
resolutions of the dE and E detectors averaged 0.3\%. A strangelet
incident on our detector system would penetrate the foil and leave
well defined signals in the two silicon detectors. For example,
$^{54}\mathrm{O}$ would deposit about 16 MeV in the dE detector and
leave the remaining 24 MeV for collection by the E detector.

Some background particles,  most abundantly knocked-out carbon ions,
survive the analyzing magnet's rigidity selection and enter our
detector system. These ions left nonzero signals in both the dE and
E silicon counters, but in no case were these signals within 10 MeV
of the expected strangelet signal on the dE vs. E plot. The
experiment was therefore free of background. The counting rates of
these carbon ions (when they appeared, which were rare cases) were
less than 10 Hz, so the dead time of the detectors was negligible.

We have searched over  a range in mass from 42 to 70 amu and found
no strangelet candidate events. The single event sensitivity limit
for strange oxygen with a given mass $A_0$ can be calculated as
\begin{equation}
s=\frac{1}{1.5 \times I \times T \times P_{+5} \times \epsilon_T(5) } \label{eq:sensitivity}
\end{equation}
where $I$ is the current  of $^{16}\mathrm{O}$  out of the ion
source. $1.5 \times I$ gives the total ion current out of lunar soil
source, considering the relative abundance of oxygen atoms in lunar
soil and negative ion forming efficiency by sputtering. $T$ is the
running time per mass setting. $P_{+5}$  is the probability of a
strangelet oxygen with given mass $A_0$, and kinetic energy 17~MeV
being stripped to a charge state of $Q=+5$. $\epsilon_T(5)$ is the
transmission efficiency of charge +5 beam through the tandem from
the source to our detectors (not including stripping probability).

The running time $T$ was  nominally two hours for each mass setting.
The current of $^{16}\mathrm{O}$, which averaged approximately
$7\times10^{13}$ particles per second, was measured with a Faraday
cup after the inflector magnet before and after each run. $P_{+5}$
is calculated for each mass from the formula given
in~\cite{Shima:1986}.  These stripping probabilities depend only on
the velocity and bare charge of the nucleus and the formula used is
a parameterization of experimental data. Because $+5$ is the most
likely stripped charge state to emerge from the carbon foil,
experimental measurements have been made at similar energies and the
interpolation via this formula introduces little uncertainty.  We
find the value to be, on average, $0.4\pm0.1$.

The dominant systematic  uncertainty in our sensitivity comes from
the determination of $\epsilon_T(5)$.  Direct measurement of
$\epsilon_T(5)$ is unfeasible because of the large uncertainty of
stripping probability $p_{+5}$ for a normal nucleus with $A_0\sim
54$ and incident energy of 17 MeV. We determined $\epsilon_T(5)$ by
measuring $\epsilon_T(Q)$ vs. stripped charge $Q$ for a variety of
mass states to determine the dependence of $\epsilon_T(Q)$ on mass
and charge. This involved improving the existing measurements for
charge state stripping for various charge states which was done in a
separate apparatus not described here. From the reproducibility of
and variation in these measurements, we estimate a systematic
uncertainty of $\pm50\%$ on $\epsilon_T(5)$.

\begin{figure}
\includegraphics[width=0.45\textwidth]{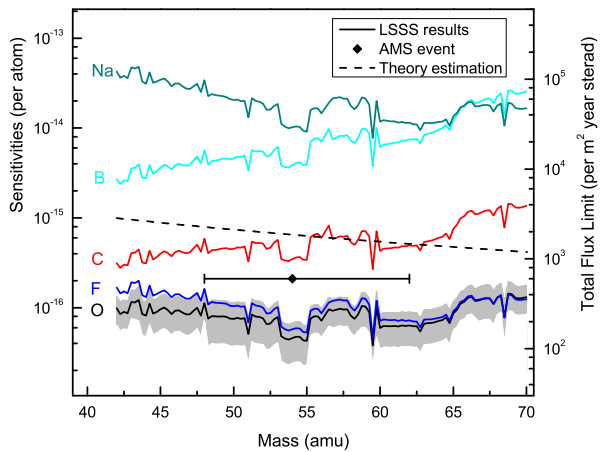}
\caption{\label{fig:lsss_results} The 95\% confidence level upper
limits of our search for different strangelet elements are shown in
solid lines, with the element names on the left. The Y axis on the
right shows the implied flux limits for cosmic ray strangelets. The
dashed line corresponds to theoretical estimation of strangelet flux
in cosmic rays at the Moon~\cite{Madsen:2004vw}. The gray area
represents the systematic uncertainties for strange oxygen search
limits.}
\end{figure}

The 95\% confidence level upper limit for strange oxygen
concentration in lunar soil is about $10^{-16}$, as shown in
Fig.~\ref{fig:lsss_results} by the black solid line. The gray area
corresponds to the systematic uncertainties of the 95\% upper
limits.

The search was optimized  in this mass range for strangelets of
nuclear charge $Z=8$ (i.e. strange oxygen), but was sensitive to
strangelets of different $Z$ values. These limits are different than
the $Z=8$ limits because nuclei of other $Z$ values will generally
have different efficiencies for producing negative ions in the
sputtering ion source and different probabilities for stripping to
$Q=+5$ in the carbon foil. When these differences are accounted for
(in the former case, by consulting ~\cite{Middleton1989} and making
measurements of source currents for various nuclei; in the latter
case, by simply using~\cite{Shima:1986}), we obtain the limits for
strange boron, carbon, fluorine, and sodium shown in
Fig.~\ref{fig:lsss_results}. This experiment is not sensitive to
strange nitrogen, neon, or magnesium at all because these elements
do not form negative ions by sputtering.

These sensitivity results  can be transformed into limits on the
strangelet flux to compare with the AMS-01 candidate and theoretical
predictions. The transformation between search sensitivity and flux
limit is determined by this lunar sample's exposure age to cosmic
rays and the distribution of strangelets versus depth that would be
expected in lunar soil. For a commonly used energy
spectrum~\cite{Madsen:2004vw}, approximately 40\% of the incident
strangelets under study here would stop in the top 7.5 cm of lunar
material~\cite{SRIMbook}, from which the lunar soil sample No. 10084
was collected. Using these numbers, the upper flux limits determined
from this search are shown by the right Y axis in
Fig.~\ref{fig:lsss_results}. Additional uncertainties introduced in
the transformation process are not considered.

Strangelet cosmic ray flux  limits can also be derived from results
of terrestrial searches for strangelets. There is more uncertainty
in flux limits derived from terrestrial searches because the final
distribution of strangelets incident on the Earth in quite
complicated~\cite{Monreal:2005dg}. However, a rough
comparison~\cite{Finch:2006pq} shows that the search reported here
is more sensitive for $Z=8$, $42<A<70$ strangelets in cosmic rays by
some 4 orders of magnitude than terrestrial searches and also gives
the best existing limit for nearby charge states.

One important result of this  search is that our limits are
inconsistent with the AMS-01 $Z=8$ candidate event at the 95\%
confidence level from 42 to 70 amu. If the AMS-02 experiment is
launched onto the ISS, it will be an almost definitive search for
cosmic ray strangelets reaching a sensitivity of 1 per (m$^2$ year
sterad) over a wide mass range (and, by extension, a very strict
test of the hypothesis of stable SQM).  In the event that AMS-02 is
further postponed or canceled, there is still room for improvement
in searches such as the one reported here.

We can extend our search to a higher mass range by simply running
more and/or altering the beam-line so that the mass acceptance is
larger (though this would of course greatly increase the expense of
the experiment). Because the search was influenced by the AMS-01
event, the covered mass region is not centered around the region
considered theoretically to be the most likely, i.e. 80 to 140 amu
(for $A<<1000$, $A=10(Z)$ for standard bag model calculations and
$A=6(Z)^{\frac{3}{2}}$ for the case when CFL is
included~\cite{Madsen:2001fu}; Ref.~\cite{Oertel} finds that CFL
strangelets may have $Z=0$ if the pairing energy is high). Also, we
could improve our limits drastically by enriching the heavy isotope
concentration in the lunar samples, though this would restrict our
sensitivity to strange oxygen only.

The existence of stable SQM remains an open question.

\begin{acknowledgments}
The authors thank P. Parker, J. Clark, A. Parikh, C. Deibel, and C.
Wrede for their great help, especially in the stripping probability
measurements. We also gratefully acknowledge the valuable
contributions of G. Lin, T. Hurteau, H. Lippincott, R. Terry, J.
Baris, T. Barker, W. Garnett, and other WNSL operators. The work is
supported by US Department of Energy under contract No.
DE-FG02-92ER-40704 and No. DE-FG02-91ER-40609, and by the Danish
Natural Science Research Council.
\end{acknowledgments}

\end{document}